\documentclass[prl,twocolumn,amsmath,amssymb]{revtex4}
\usepackage{graphicx}
\usepackage{indentfirst}
\usepackage{psfrag}
\usepackage{epsfig}
\usepackage{amsmath}
\usepackage{amssymb}
\usepackage{bm}
\usepackage{mathrsfs}
\newcommand{\be}{\begin{eqnarray}}
\newcommand{\ee}{\end{eqnarray}}

\begin{document}

\title{Comment on "Poynting vector, orbital and spin momentum and angular momentum versus optical
force and torque on arbitrary particle in generic optical fields" } 
\author{Manuel Nieto-Vesperinas }
\affiliation{Instituto de Ciencia de Materiales de Madrid, Consejo Superior de
Investigaciones Cient\'{i}ficas\\
 Campus de Cantoblanco, Madrid 28049, Spain.\\ www.icmm.csic.es/mnv; 
mnieto@icmm.csic.es }



\begin{abstract}
We criticize the originality or correctness of some of the ideas and results recently reported by Ng et al. in arXiv:1511.08546.
\end{abstract}
\maketitle

A recent paper \cite{ng}   addresses the optical force and torque on a general particle illuminated by what the authors call a generic monochromatic free-space optical field ${\bf E}^{(i) }$,  ${\bf B}^{(i) }$. Among its results pertaining to a multipole expansion of these quantities, it is claimed to show, (next we quote and comment):

{\bf -}  "The optical torque originates not only from the spin angular momentum (AM) but also from the orbital AM, clarifying in a generic case the long-standing controversy about whether the orbital AM can induce a spinning torque... The optical torque is brought about by the total AM, both the orbital and the spin AM".

$\bullet$ The authors do not mention, nevertheless, that such result was already obtained in \cite{MNV2015_1,MNV2015_2} on which   \cite {ng} clearly stands, (also, no wonder \cite {ng} appears shortly after \cite{MNV2015_1,MNV2015_2} are published). In particular, it was shown and extensively discussed in \cite{MNV2015_2} that this influence of the orbital AM on the optical torque is a consequence of its continuity equation, adding to that from the spin AM as a manifestation of the spin-orbit interaction.

{\bf -}  "in contrast with previous studies on the optical torque," (in which they include \cite{MNV2015_1,MNV2015_2}) "which  are confined to the small particle limit and from which no universal conclusion can be reached", \cite {ng} establishes "general equations", [their Eqs. (3a) and (3b)],  that "provide a more transparent and general physical picture for understanding the optical force and torque as mechanical manifestations of the linear optical momentum and the AM".

$\bullet$ However, nowhere  the authors of \cite{ng} explain what phenomena, different from  those already derived in previous works for  wide sense dipolar particles (this involves bodies much larger than those of the small particle limit \cite{MNV2015_1,MNV2015_2}) and duplicated in  \cite{ng},  are obtained from those authors multipole expansion.

Such Eqs.(3a) and (3b) are proven through the use of Jones'  lemma based on the principle of the stationary phase and on a generalization made by the authors. To do this, they decompose the fields into incident and scattered, and introduce them, as usual, into the flow of Maxwell's stress tensor (ST) evaluated in the far-zone; then they state: "The mixed and scattering terms yield, respectively, the extinction and recoil forces", a fact that the authors quote derived in \cite{chen1}, (and repeatedly reported in Ref. \cite{ chen2}  that they also cite).

$\bullet$  However,  we should remark that, as far as dipolar particles in  the wide sense are concerned,  both the stationary phase method of ST flow evaluation in the far-zone, and its above mentioned  result  in Ref.\cite{chen1},  actually   duplicated the  work already reported in \cite{opex2010}, where in addition to the extinction and recoil force components,   these terms were shown  to describe  the pure electric, and/or magnetic,  and the electric-magnetic interaction components of the optical force from fields represented   by a angular spectrum of plane waves. 

In this regard we criticize  that the authors of \cite {ng} call "generic" those fields that,  like in \cite{MNV2015_1,MNV2015_2, opex2010}, are   expressed by an angular spectrum of plane waves. It  is well-known that, although such representation characterizes a wide variety of optical fields, it does not correspond to a "generic" wavefield, (see \cite{,MNV2015_2} and Refs. [65] and [66] therein).

{\bf -} After obtaining a multipole expansion of the optical torque, the authors of \cite {ng}  particularize it to dipolar particles, so getting a time-averaged torque expressed as the sum of two terms, which apart from constant factors which depend on the system of units, are respectively proportional to:

\be
\frac{1}{2  } \Re [ ({\bf p}\times {\bf E}^{(i) *} ) + ( {\bf m}\times{\bf B}^{(i) *} )] ,  \label{torq2}
\ee

and

\be
-\frac{k^3}{3}  [\frac{1}{\epsilon}\Im( {\bf p}^{*}  \times {\bf p}) + \mu \Im ( {\bf m}^{*}  \times {\bf m}) ] \label{ttss1},
\ee
which the authors recognize as the extinction and the recoil term, respectively. $\Re$ and $\Im$ denote real and imaginary parts, whereas $\epsilon$ and $\mu$ represent the permittivity and permeability of the host medium, respectively. $k=\sqrt{\epsilon \mu}\omega/c$.

Then they state: "The recoil part has been long missing in many previous studies  even in the dipole limit. Actually, it is the recoil torque that cancels out the extinction torque on any non-absorbing spherical particle" (quoting their Ref. [50])..."Besides the extinction optical torque originating from the interception of the incident photons as a simple extension to the static case, the theory includes the long missing recoil torque that stems from the interference of
reradiation by the multipoles of the same type and order induced on the particle in optical field".

$\bullet$ We remark, however, that although  Eqs.(\ref{torq2}) and (\ref{ttss1})  are consistent with a previous report by the authors on systems of larger particles, quoted as their  Ref.[50], they do not acknowledge  in \cite{ng} that such equations do not constitute a new result  of  \cite{ng} since  (\ref{torq2}) and (\ref{ttss1}) were already established in our recent papers \cite{MNV2015_1,MNV2015_2}, where we already pointed out  the missing recoil torque (\ref{ttss1}) in previous studies dealing with dipolar particles, as a consequence of an extensive use through the years of the static approximation.

In addition, contrary to what it is stated in \cite{ng}, we believe that the expression (\ref{torq2})  for the extinction torque is valid for incident plane wave illumination, or for a static dipole, but not for an arbitrary wave. For the latter, as a consequence of the conservation of spin and orbital AMs, and as shown in \cite{MNV2015_1,MNV2015_2}, the extinction torque contains the additional term proportional to: 
\be
\Im\{\frac{1}{\epsilon}({\bf p} \cdot\nabla){\bf B}^{(i) *}-\mu  ({\bf m} \cdot\nabla){\bf E}^{(i) *} \}; \label{shape}
\ee
which, vanishing for a plane wave, accounts for effects due to the spatial  structure and polarization of the incident wave. 

{\bf -} "Actually, it is the recoil torque that cancels out the extinction torque on any non-absorbing spherical particle", and the authors quote their Ref.[50].

$\bullet$ We argue, however, that this latter result, which  involves the conservation of energy and that was previously shown on employing Mie's theory in \cite{marston} applied to a sphere of arbitrary size illuminated by a plane wave, had already been generalized again in  \cite{MNV2015_1,MNV2015_2}  for a field expressible by an angular spectrum representation;  [see also a detailed discussion for optical fields in Section X of \cite{MNV2015_2} where it is shown that the term (\ref{shape}) remains after that cancellation].

 It is frustrating that with the rise of research papers, which makes it increasingly difficult for  reviewers to identify previous work  in the scientific literature, there is a growing incidence of  reported results which are actually duplications of earlier studies without proper acknowledgement, whether this is unintentional  or not.

\end{document}